\documentclass[aip,reprint]{revtex4-1}
\usepackage{graphicx}
\usepackage{fancyhdr} 
\usepackage{xcolor}
\draft 
\newcommand{\rev}[1]{{\color[rgb]{0 0 0}{#1}}}

\fancypagestyle{firstpage}{%
  \cfoot{\footnotesize This article may be downloaded for personal use only. Any other use requires prior permission of the author and AIP Publishing. This article appeared in Appl. Phys. Lett. 119, 263504 (2021) and may be found at https://doi.org/10.1063/5.0069116.}
}
\begin{document}


\title[Y-Flash memristor model]{Physical based compact model of Y-Flash memristor for neuromorphic computation} 



\author{Wei Wang}
\email{E-mail: wei.wang@campus.technion.ac.il}
\author{Loai Danial}
\email[Current address: Intel Corporation, IDC, Haifa]{}
\author{Eric Herbelin} 
\author{Barak Hoffer} 
\author{Batel Oved} 
\author{Tzofnat Greenberg-Toledo}
\affiliation{The Andrew and Erna Viterbi Faculty of Electrical and Computer Engineering, Technion--Israel Institute of Technology, Haifa 3200003, Israel}
\author{Evgeny Pikhay}
\author{Yakov Roizin}
\affiliation{Tower Semiconductor, Migdal HaEmek 2310502, Israel}
\author{Shahar Kvatinsky}
 \email[]{Authors to whom correspondence should be addressed: shahar@ee.technion.ac.il}
\affiliation{The Andrew and Erna Viterbi Faculty of Electrical and Computer Engineering, Technion--Israel Institute of Technology, Haifa 3200003, Israel}


\date{\today}

\begin{abstract}
Y-Flash memristors utilize the mature technology of single polysilicon floating gate non-volatile memories (NVM). It can be operated in a two-terminal configuration similar to the other emerging memristive  devices, \textit{i.e.}, resistive random-access memory (RRAM), phase-change memory (PCM), \textit{etc.} Fabricated in production  complementary metal-oxide-semiconductor (CMOS) technology, Y-Flash memristors  allow excellent  reproducibility  reflected  in high neuromorphic products yields. Working in the subthreshold region, the device can be programmed to a large number of  fine-tuned intermediate states in an analog fashion and allows low readout currents (1 nA $\sim$ 5 $\mu$A). However, currently, there are no accurate models to describe the dynamic switching in this type of memristive device  and  account for multiple operational configurations. In this paper, we provide a physical-based compact model that describes Y-Flash memristor performance both in DC and AC regimes, and  consistently  describes the  dynamic program and erase operations. The model is integrated into the commercial circuit design tools and is ready to be used in applications related to neuromorphic computation.
\end{abstract}


\maketitle 

\thispagestyle{firstpage}

Emerging non-volatile memory devices (NVMs), such as resistive random-access memory (RRAM)\cite{Yang2013}, phase-change memory (PCM)\cite{Zhang2019}, ferroelectric random-access memory (FeRAM)\cite{Max2020}, and electrolyte-gate transistors\cite{Li2020,Burgt2017}, have tunable conductance which can faithfully emulate the plasticity of synaptic connections in an artificial neural network and thus are promising for the large-scale implementation of neuromorphic computations\cite{Tang2019,Burr2017,Ielmini2018}. However, these emerging technologies are still not mature enough for production-worthy neuromorphic systems, mostly because of low yields and high non-uniformity of the memristive devices\cite{Gokmen2016,Chen2017,Dittmann2019,Berggren2021}. Conversely, NVM floating gate (FG) devices are fabricated in standard  complementary metal-oxide-semiconductor (CMOS) processes. They have also been proposed as candidates for artificial synapses~\cite{Ziegler2012,Malavena2019}. As its name suggests, the FG device has a floating gate isolated from its other terminals. The FG can be charged or discharged; thus the threshold of the corresponding transistor can be adjusted to the desired level. 

Y-Flash devices have a simpler configuration compared with conventional FG devices: the control gate is merged with drain, thus largely decreasing the device footprint, while program, erase, and readout operations are possible in two-terminal configuration~\cite{Roizin2016}. Arranged in a crossbar array, the Y-Flash devices are suitable for accelerating the frequent and expensive vector-matrix multiplications (VMMs) acting as a synaptic array for multiple neuromorphic applications\cite{Danial2019}. The Y-Flash memristor works in an analog fashion and can be operated in the subthreshold range, achieving a large number of tunable conductance states and low readout currents. \rev{Compared with other NVMs, the Y-Flash memristor shows a combination of several advantages, including fully CMOS process compatibility, low cycle-to-cycle variations, high yield, low power consumption, analog conductance tunability, self-selection, and high retention time.}

The success of a neuromorphic computation system needs the co-design of the synaptic array (\textit{e.g.}, Y-Flash device array) and the CMOS-based peripheral circuit, \textit{i.e.}, artificial neurons\cite{Wang2020}. Thus, a compact model which can accurately simulate the synaptic characteristics of the Y-Flash device to enable the full design flow of the neuromorphic computation system is needed. Previously, we reported a preliminary compact model which could capture the basic program, erase, and readout operations for a fixed configuration and operation voltage based on empirical equations\cite{Danial2020}. However, the accurate I-V characteristics, as well as the program and erase operations at different configurations and voltage biases, could not be simulated by the preliminary model (see Supplementary Materials Table S1). The  preliminary model did  not support transient analyses either, since the effects of parasitic capacitors, as well as other parasitic effects, were not considered. 

In this paper, we provide a comprehensive model with in-depth physical considerations related to the program/erase mechanism and the account for the parasitic effects. The FG voltage is determined by the analysis of the coupled capacitor network, and the I-V characteristics consider the behaviors both in the subthreshold and above-threshold regions. The program operation is modeled by the lucky hot electron injection, and the erase operation is described as a  combination of band-to-band tunneling (BBT) of holes, their acceleration in the lateral field, and tunneling through the dielectric layer. The simulation results show excellent agreement with the experimental data. The model is written in Verilog-A language and is integrated into a commercial circuit design simulator to enable the co-design of a neuromorphic system with Y-Flash memristors and CMOS peripheral circuits.

\begin{figure*}
\includegraphics[width=0.9\textwidth]{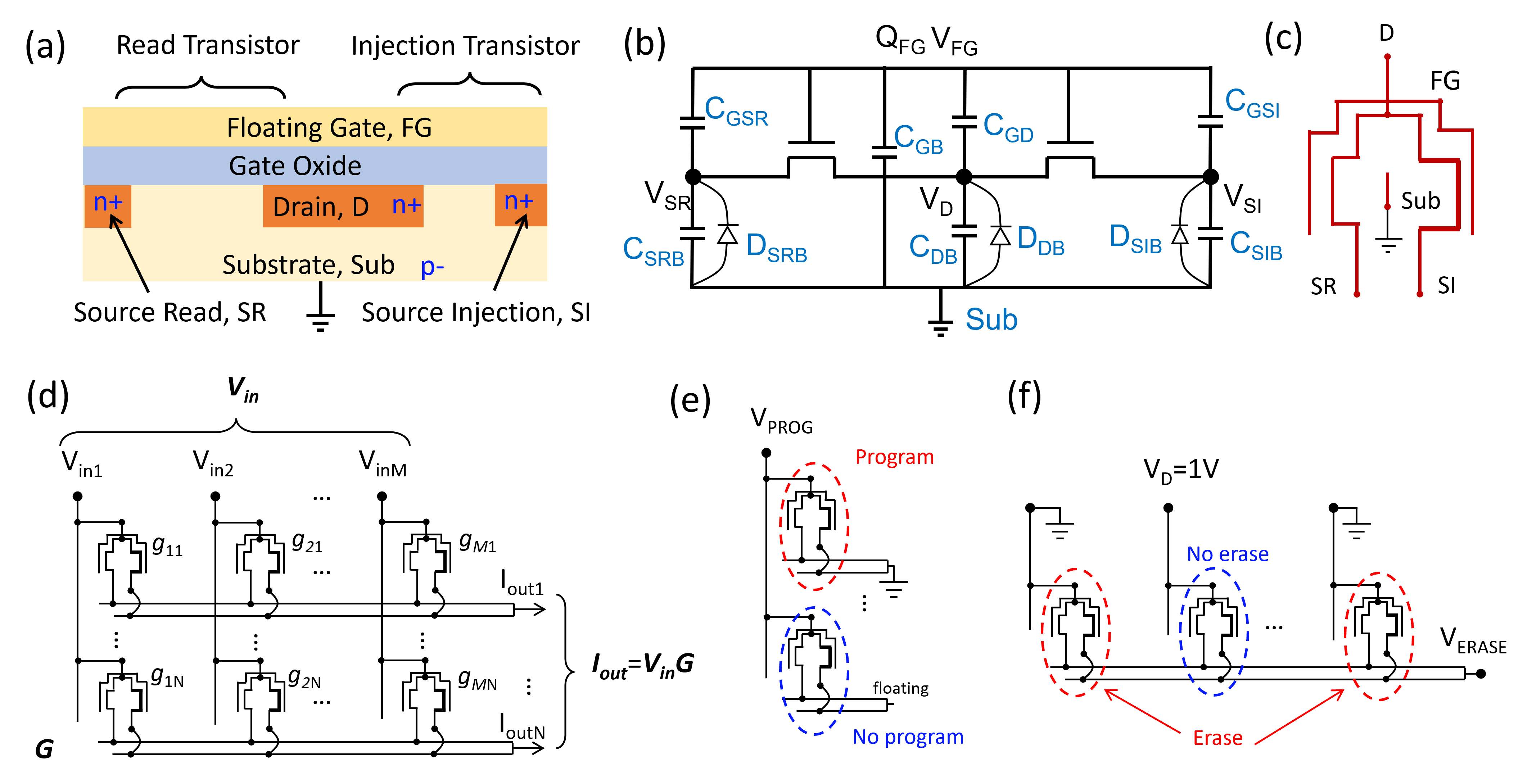}%
\caption{\label{fig1}(a) Schematic of the Y-Flash structure; (b) equivalent circuit accounting for parasitic capacitors, transistors, and p-n diodes; (c) the symbol of the Y-Flash device with external terminals; (d) Y-Flash array for VMM; (e) programming of the devices in an array; (f) erasing of the devices in an array. }%
\end{figure*}

The structure of the Y-Flash device is shown in Fig.~\ref{fig1}a, which consists of two transistors, \textit{i.e.}, a read transistor and an injection transistor, with a common FG and a common drain (D). \rev{The read transistor (channel length=6$\mu m$) has a lower threshold voltage by a different channel doping optimized for the readout, while the injection transistor has a shorter channel (3$\mu m$) to enhance the hot carrier injection.} The capacitance of the common D is larger than the source of read transistor (Source Read, SR) and the source of injection transistor (Source Injection, SI) to couple the FG more on the D terminal. Fig.~\ref{fig1}b and Fig.~\ref{fig1}c show the equivalent circuit, accounting for parasitic capacitors and p-n junctions between substrate and terminals, and the symbol of the Y-Flash device, respectively. The FG is isolated from the  external terminals and can store charge ($Q_{FG}$ in Fig. 1b), and the substrate is always grounded. The device was fabricated using standard CMOS process flow (Tower Semiconductor 180nm) without any additional masks. More details can be found in our previous publication\cite{Danial2019}. 

Although the Y-Flash device has three external terminals, the two sources, SR and SI, can be shortened (No. 1, 3, 6 in Table I), such that the device acts as a two-terminal memristor. The two-terminal Y-Flash  configuration enables easier addressing by the peripheral circuits, while the three-terminal configuration enables more flexible and efficient read, program, and erase operations (No. 2, 4, 7 in Table I). Read operation is conducted by applying a low read voltage ($V_R<2.5V$) on the D of the device (No. 1, 2 in Table I). Similar to other memristive arrays, VMM can be performed directly in a Y-Flash device array leveraging Ohm’s law and Kirchhoff’s current law (Fig. 1d). At a high drain voltage ($V_{P}>4V$) and with one or two of the source terminals grounded, there will be hot electron injection into the FG, and the threshold voltage of both transistors will be increased, \textit{i.e.}, corresponding to device programming. The program will be disabled if both the SR and SI are floating (No. 5 in Table I), which can be used to deselect the devices that are not intended to be programmed in an array (Fig. 1e). The erase operation (decrease of the threshold voltage, or discharge of the floating gate) can be achieved by applying a high voltage ($V_{E}>7V$) on the SI terminal and leaving the drain terminal floating \rev{or grounded} (No. 6, 7 in Table I). The devices within the same row which, however, are not intended to be erased, can be deselected by applying a certain  voltage on the drain (No. 8 in Table I and Fig. 1f).

\begin{table}
\caption{\label{tab1}Operation configurations of the Y-Flash devices. }
\begin{ruledtabular}
\begin{tabular}{ccccc}
No. & Mode & D & SR & SI\\
\hline
1 & Read & $V_R<2.5V$ & GND & GND \\
2 &      & $V_R<2.5V$ & GND & Floating \\
3 & Program & $V_{P}>4V$ & GND & GND \\
4 &         & $V_{P}>4V$ & Floating & GND \\
5 &         & $V_{P}>4V$ & Floating & Floating \\
6 & Erase & \rev{Floating/GND} & $V_{E}>7V$ & $V_{E}>7V$ \\
7 &       & \rev{Floating/GND} & Floating/GND & $V_{E}>7V$ \\
8 &       & $1V<V_D<2V$ & Floating/GND & $V_{E}>7V$ \\
\end{tabular}
\end{ruledtabular}
\end{table}

The I-V characteristics of the device can be modeled by combining the characteristics  of the two transistors with a certain charge in the FG. The potential on the FG is not directly given, and in several operation modes, as indicated in Table I, some terminals are floating. These unknown potentials can be obtained by calculating voltages in the circuit of parasitic capacitors, p-n junctions, and equivalent resistors of the transistor channels. When all the three terminals are externally connected to voltage sources, the FG potential can be obtained by solving the equation of the fixed total charge in the floating gate,
\begin{eqnarray}
Q_{FG}=&&C_{GSR} (V_{FG}-V_{SR})+C_{GD} (V_{FG}-V_D )\nonumber\\
       &&+C_{GSI} (V_{FG}-V_{SI} )+C_{GB} V_{FG},
\end{eqnarray}
where $C_{GSR}$, $C_{GD}$, $C_{GSI}$, and $C_{GB}$ are the capacitors between FG and SR, FG and D, FG and SI, FG and substrate, respectively, and, $V_{SR}$, $V_D$, and $V_{SI}$ are the voltages on SR, D, and SI, respectively. When one or two of the three external terminals are floating, their potentials are obtained from the circuit shown in Fig. 1b. Note, that the solution corresponding to  the circuit in Fig. 1b can be obtained  by the circuit simulator directly when the circuit is written in hardware description languages. 

To model the wide range of operating voltages of the Y-Flash device, the I-V characteristics  both in subthreshold and above-threshold regions should be accurately described. The source-drain current in the subthreshold region ($V_{FG}<V_{TH}$) for both transistors can be written as 
\begin{equation}
    I_{DS,sub}=I_{S0} e^{q\frac{V_{FG}-V_{TH}}{nkT}} (1-e^{-q \frac{V_D-V_S}{kT}}),
\end{equation}
where $I_{S0}$ is a pre-factor, $V_{TH}$ is the intrinsic threshold voltage, $q$ is the elementary charge, $n$ is the ideality factor, $k$ is the Boltzmann constant, and $T$ is the temperature. Note that the parameters $I_{S0}$ and $V_{TH}$ are different for the two transistors, and $V_S$ is the voltage on the source terminal which should be replaced by $V_{SR}$ and $V_{SI}$ for the read transistor and injection transistor, respectively. Above threshold ($V_{FG}>V_{TH}$), the source-drain current can be written as
\begin{widetext}
\begin{equation}
    I_{DS,ab}=\Big\{
    \begin{array}{ll}
    \frac{K}{2} (V_{FG}-V_{TH} )^2, & V_{FG}-V_{TH}<V_{DS}  \\
    K(V_{FG}-V_{TH}-\frac{V_{DS}}{2})V_{DS}, & V_{FG}-V_{TH}>V_{DS}\\
    \end{array}
\end{equation}
\end{widetext}
where $K=\frac{W}{L}\mu C_{ox}$, $W$ is the channel width, $L$ is the channel length, $\mu$ is the mobility, and $C_{ox}$ is the gate oxide capacitance. \rev{Note that the $V_{FG}$ is the function of the D voltage, as shown in the band diagram for reading operation in Supplementary Fig. S1a-b.} The current should be continuous between the two regions. Thus, the two currents in Eq. (2) and Eq. (3) are extended to the full voltage range and combined by a smooth function,
\begin{equation}
I_{DS}=(\frac{1}{I_{DS, sub}^m}+\frac{1}{I_{DS,ab}^m})^{-\frac{1}{m}},
\end{equation}
where $m$ is a smooth factor ($m=1$ is used in our model). 

\begin{figure*}
\includegraphics[width=0.7\textwidth]{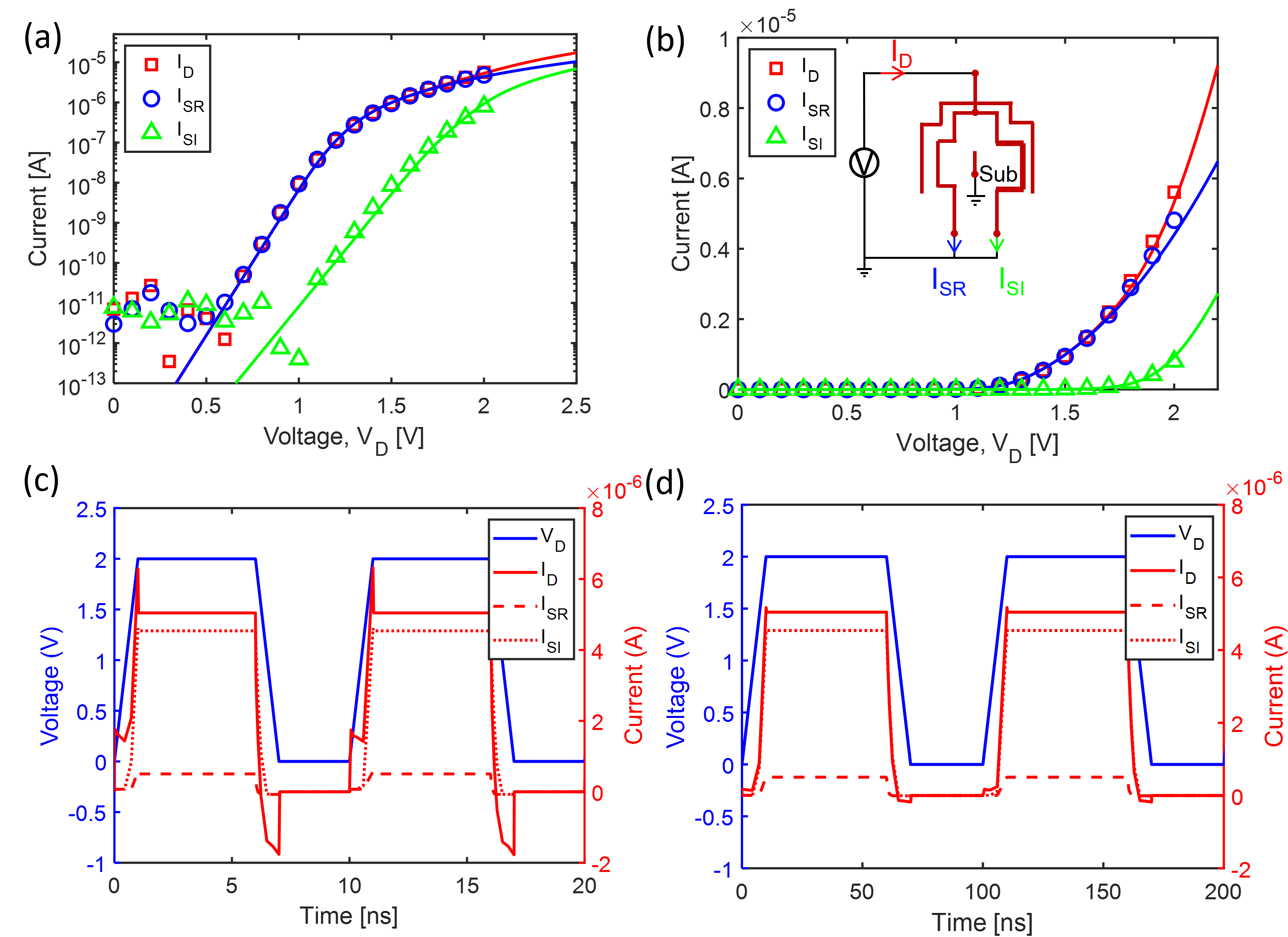}%
\caption{\label{fig2}Comparison of the measured and modeled DC I-V characteristics Y-Flash device in read mode for a pristine (non-programmed) device ($Q_{FG}=0$): (a) in logarithmic scale; (b) in linear scale (inset: schematic of the read operation). Pulse reading of the model implemented in circuit simulator: (c) pulse width 5 ns, rise time 1 ns; (d) pulse width 50 ns, rise time 10 ns.}%
\end{figure*}

Fig. 2a and Fig. 2b show the comparison between the measured DC read currents of the device and fitting lines by the model, in logarithmic and linear scales, respectively. Accounting for the parasitic capacitors in the model, the overshoot effects in the pulse measurements can be also simulated in transient mode (Fig. 2c and 2d). The device states can also be read with the SI or SR being floating (see Supplementary Materials Fig. S2 for the model results of additional read configurations). 

\rev{Note that it is only possible to operate the device with positive voltages since the substrate is always grounded and there are p-n junctions between the p-type substrate and the n+ type ohmic drain/source contacts. Negative reading of the device, however, can be performed by grounding the D and applying a positive voltage on the SR and SI, as shown in Supplementary Fig. S3. The negative reading shows ultra-low current since the floating gate is coupled to the D and the transistors are turned off. The low negative reading current enables the self-selection and low sneak path current of the devices in an array.}

\begin{figure*}
\includegraphics[width=0.9\textwidth]{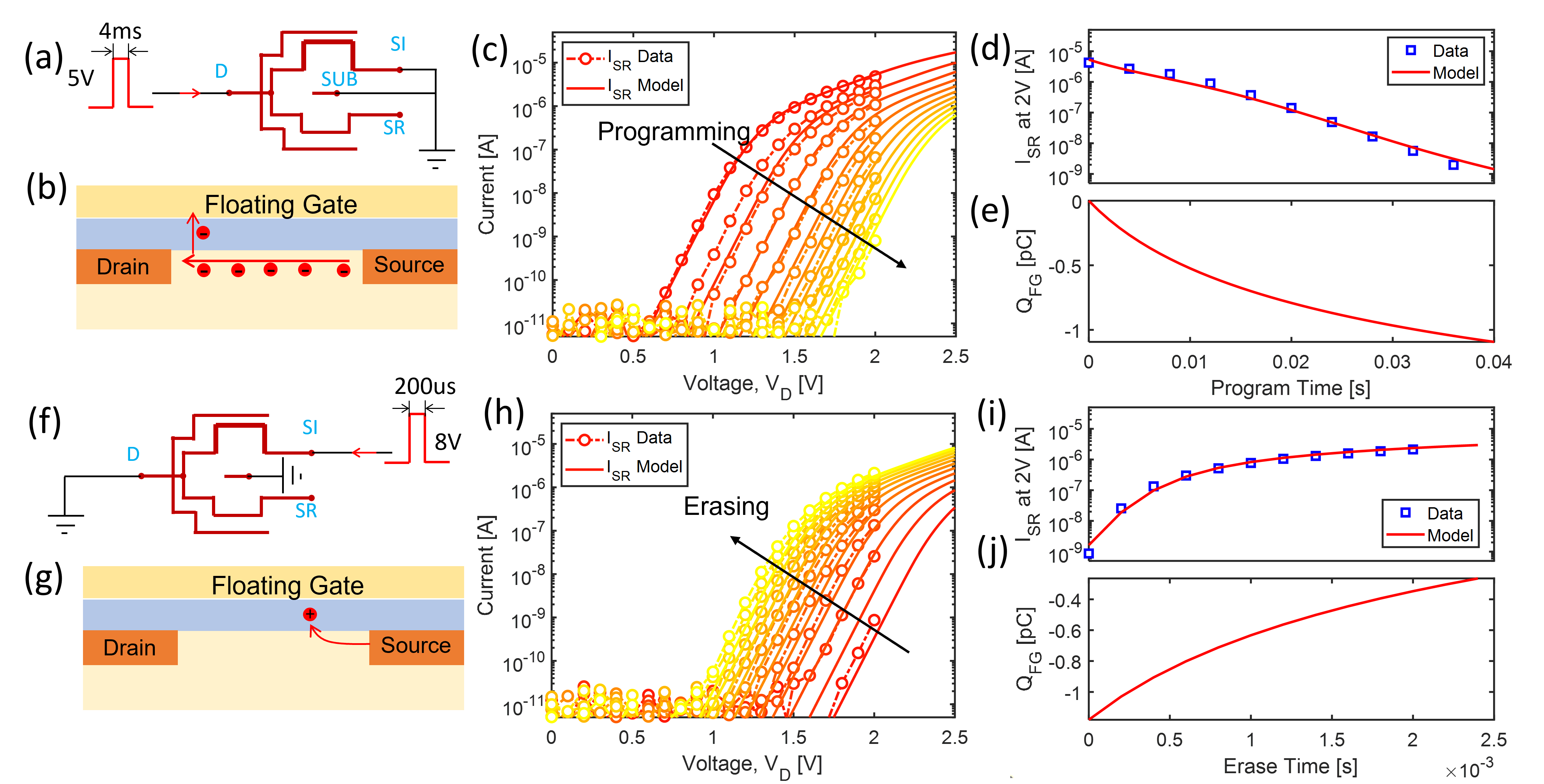}%
\caption{\label{fig3}(a) Schematic of the program operation; (b) mechanism of hot carrier injection into the FG in program operation; (c) readout I-V characteristics for subsequent programming pulses; (d) readout current at 2V as a function of accumulated program time; (e) FG charge as a function of the accumulated program time; (f) schematic of the erase operation; (g) mechanisms of FG discharge in  the erase operation; (h)  readout of the device for subsequent erase pulses; (i) readout current at 2V as the function of accumulated erase time; (j) FG charge as a function of the accumulated erase time.}%
\end{figure*}

In the following measurements and simulations, we only grounded the SI terminal in the program mode (Fig. 3a). The program operation (injections of the electron to the FG) is conducted by the lucky hot carriers. We use a simplified model\cite{Diorio1996} where the maximum injection into the FG is at the point where the FG potential is equal to the channel potential\cite{Diorio1996,SimonTam1984} (Fig. 3b \rev{and Supplementary Fig. S1c-d})
\begin{equation}
I_{G,inj}=-I_{DS} P_{0} e^{-\frac{V_\alpha}{V_{FG}}}, 						\end{equation}
where $I_{DS}$ is the source-drain current determined by Eq. (4), $P_{0}$ is the probability of the hot electrons being emitted to the FG, and $V_\alpha$ is a fitting parameter. \rev{Note that Fowler-Nordheim tunneling usually happens in a higher voltage range ($> 10V$)\cite{Pavan1997} thus is not considered in the current model.}

To perform  erase, we apply voltage on the SI terminals, and leave the SR terminal floating, as shown in Fig. 3f. The erase is performed by the injection of holes to the FG, which were generated by BBT in the source/channel junctions and accelerated in the lateral field (Fig. 3g \rev{and Supplementary Fig. S3e-f})\cite{Yoshikawa1990,Ielmini2006}
\begin{equation}
I_{G,tunnel}=\xi(V_{FG}-V_{bi} )^2 e^{-\frac{\beta}{V_{FG}-V_{bi}}}, 	
\end{equation}
where $\beta$ and $\xi$ are fitting parameters reflecting hole injection efficiency into FG, and $V_{bi}$ is the potential at the interface with silicon at the point where injection of holes takes place.

The change of the charge on the FG can be modeled by combining the two contributes of gate current,
\begin{equation}
\frac{dQ_{FG}}{dt}=I_{G,inj}+I_{G,tunnel},									
\end{equation}
where $t$ is the real-time. \rev{We assume that the as-fabricated device initially has no charge on the floating gate.}

\begin{table}
\caption{\label{tab2}Parameters and their values in the model. }
\begin{ruledtabular}
\begin{tabular}{ccc}
Parameters & Read Transistor & Injection Transistor\\
\hline
$C_{GD}$ & \multicolumn{2}{c}{1.0 fF}\\
$C_{GB}$ & \multicolumn{2}{c}{0.24 fF}\\
$C_{DB}$ & \multicolumn{2}{c}{0.64 fF}\\
$C_{GSR}, C_{GSI}$ & 49 aF  & 48 aF\\
$C_{SRB}, C_{SIB}$ & 32 aF  & 32 aF\\
$V_{TH}$ & 0.82 V  & 1.34 V\\
$I_{S0}$ & 40 nA   & 80 nA \\
$K$      & $1.9\times10^{-5}A/V^2$ & $3.8\times10^{-5}A/V^2$ \\
$n$      & 1.7   & 2.21 \\
$P_{0}$  & -     & $3.8\times10^{-5}$ \\
$V_{\alpha}$ & - & 20 V\\
$\sigma_{V_{\alpha}}$ & - & 0.8 V\\
$\beta$     & - & 10 V\\
$\sigma_\beta$  & - & 0.8 V\\
$V_{bi}$    & - & 5.5 V\\
$\xi$       & - & $3.9\times10^{-12} A/V^2$\\
\end{tabular}
\end{ruledtabular}
\end{table}

To enable precise control of the states of the Y-Flash device, the voltage pulses with precise time width are employed to program and erase the devices (Fig. 3a and 3f). The program pulse in Fig. 3a has an amplitude of 5V, a pulse width of 4 ms, and the rise and fall times of 10 us. After the application of each program pulse, a sweep voltage from 0V to 2V with the configuration shown in Fig. 2b inset was applied to measure the state of the device. \rev{Fig. 3c shows the readout I-V curves before and after each of the consecutive program pulses (only currents on the SR are plotted for simplicity).} The circles are experimentally measured data, and the lines are simulation results. The readout currents at 2V are collected to dedicate the states of the devices as a function of accumulated program time in Fig. 3d, comparing the experimental data and simulation results. Fig. 3e shows the charge on the FG extracted from the simulation, as a function of accumulated program time. Note that it takes 9 pulses to program the device from the low resistance state (LRS) of readout current being 5 $\mu$A to high resistance (HRS) states of the readout current approximating 1 nA, thus ten discrete conductance states are measured. However, the program pulse width can be lowered to achieve more conductance states. More than 1000 analog conductance states could be obtained when using program pulses of 10 us (Supplementary Materials Fig. S4a). The longer pulses used here are mainly to accelerate the measurement of a full program cycle from LRS to HRS.

Similarly, the erase operation is performed by alternatively applying an erase pulse, as illustrated in Fig. 3c, and performing the readout operation. Voltage pulses with an amplitude of 8V and a width of 200 us are used for the erase operation. Fig. 3h shows the readout I-V curves during the erase operation. The read currents at 2V are presented in Fig. 3i to illustrate the states of the devices as a function of the accumulated erase time. Fig. 3j shows the charge on the FG  as a function of the erase time. Similar to the program operation, more analog conductance states can be achieved by lowering the pulse width of the erase operation (Supplementary Materials Fig. S4b).

\rev{The reading current evolution curves in Fig. \ref{fig3}d and \ref{fig3}i mimic the long-term depression (LTD) and long-term potentiation (LTP) of a biological synapse, respectively. By properly engineering the pre-synaptic pulse in the D and the post-synaptic pulse in the SI, the spike-timing-dependent plasticity (STDP) can also be emulated by the device\cite{Danial2019}. }

As indicated in Table I, the Y-Flash device can be programmed and erased in multiple configurations (for instance, SR is floating or shorted to SI), all of which can be simulated by the current model (See Supplementary Materials Fig. S5-S6 for more program/erase modes). \rev{It can be seen clearly that the three-terminal configuration is more efficient than the two-terminal one. This is largely due to the lowering of FG to channel voltage by the coupling of the FG to the SR when SR is shortened to SI.} Varing the amplitude of the program or erase pulses, the program and erase operations will result in different program and erase curves as a function of program time and erase time, respectively, which can also be accurately captured by the model (see Supplementary Materials Fig. S7). The complete model has been implemented in Verilog-A (Supplementary Materials Table S3) and enables the circuit design and simulation in the Cadence Virtuoso tool (Supplementary Materials Fig. S8, all model equations are summarized in Supplementary Table S2).

\begin{figure}
\includegraphics[width=\columnwidth]{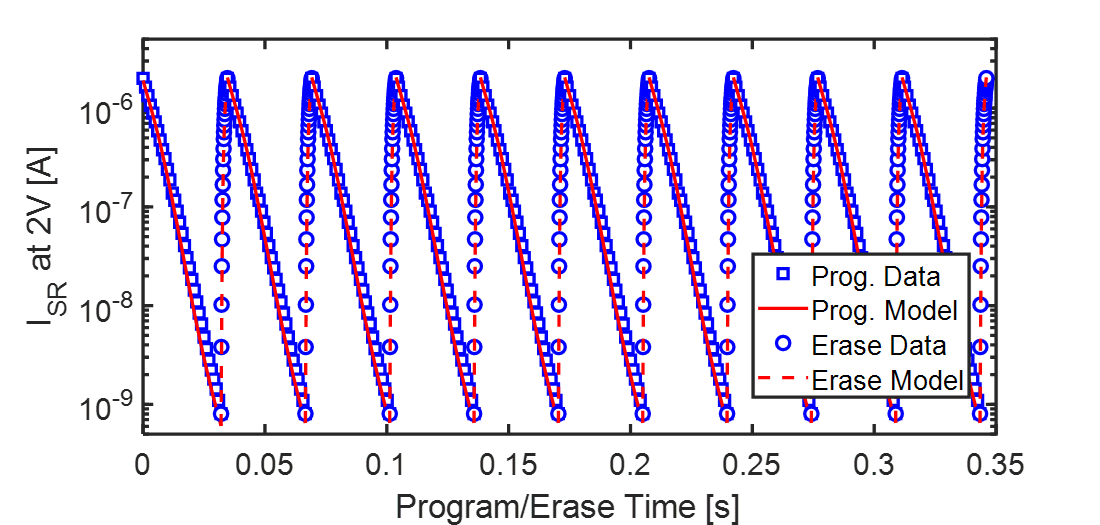}%
\caption{\label{fig4}Cycling of the program and erase operations showing excellent cycle-to-cycle uniformity. }%
\end{figure}

To illustrate the cycling performance  of the Y-Flash memristor, the program/erase cycling test was conducted. We program the device ($V_P=5V$) from the high conductive states ($I_{SR}>2\mu A$ at 2V) to low conductive states ($I_{SR}<1 nA$ at 2V) by a set of program pulses, then erase the device ($V_E=8V$) backward. After 100 cycles, no performance degradation was detected. Fig. 4 shows experimental data for ten cycles, together with the corresponding simulation results by the compact model. 

\begin{figure}
\includegraphics[width=\columnwidth]{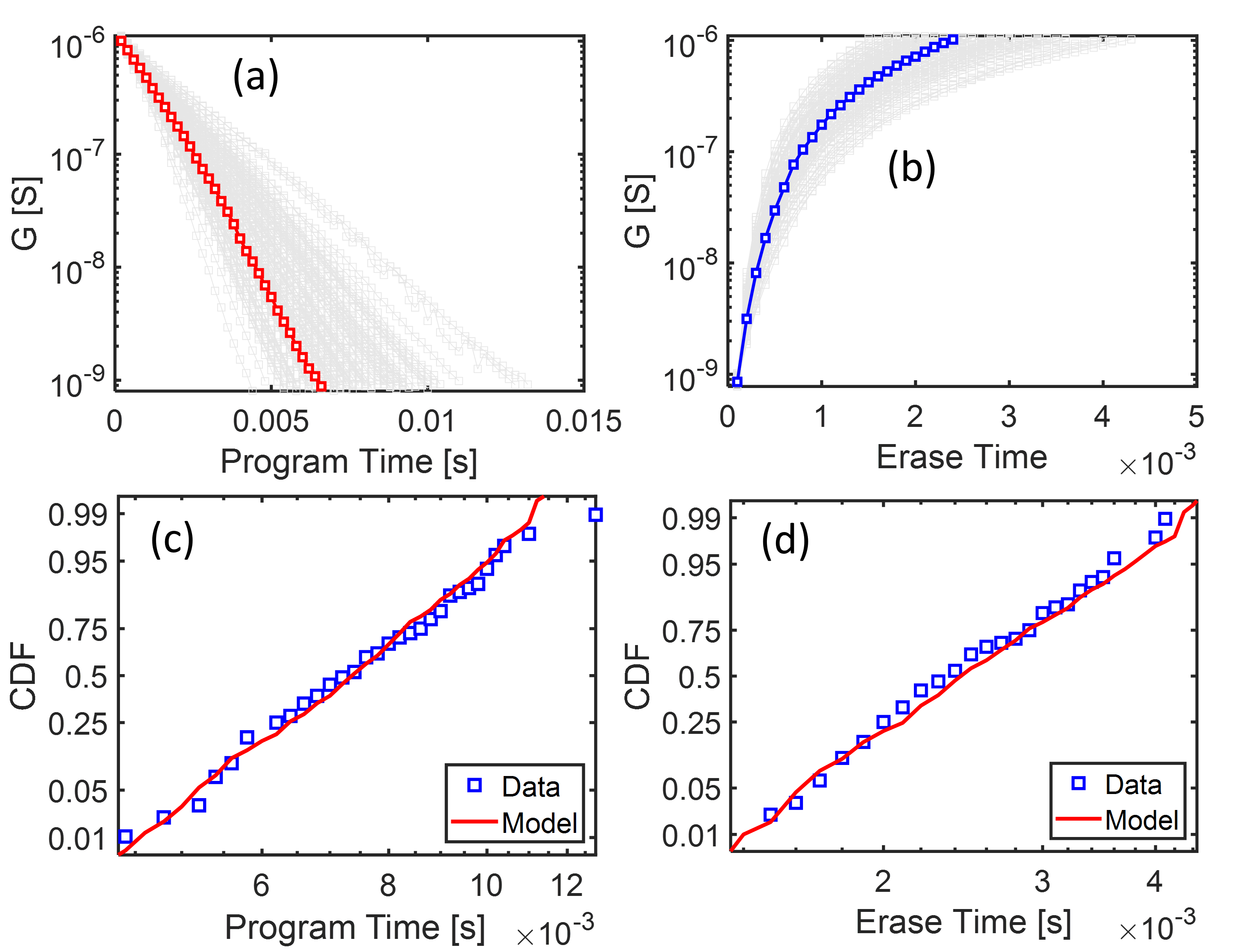}%
\caption{\rev{\label{fig5}Device-to-device variations and modeling. Device-to-device variations for (a) the pulsed program operations ($V_P=5V$, pulse width 200 $\mu$s), and (b) the pulsed erase operations ($V_E=8V$, pulse width 100 $\mu$s); Statistical total (c) program time and (d) erase time from the experiments and model results.}}%
\end{figure}

\rev{We further investigated the device-to-device variations of the Y-Flash memristor. We measured 96 devices in an array (12-by-8) of the devices. The conductance traces during the program and erase operations as a function of program time and erase time are shown in Fig. \ref{fig5}a and \ref{fig5}b, respectively. The colored lines are typical traces, and the gray lines are all the traces of 96 devices. There do exist device-to-device variations, however, all the devices work normally, proving the high yield of the fabricated devices. The statistical results of the total program time and total erase time are shown in Fig. \ref{fig5}c and \ref{fig5}d, respectively, both showing log-normal distributions. The device-to-device variation can be modeled by adding a Gaussian-type variation to the parameters in the exponents of the gate currents for the program and erase operations in Eq. (6) and Eq. (7), that is, }
\begin{equation}
    \rev{V_{\alpha,d2d} \in N(V_\alpha,\sigma_{V_\alpha}^2) }
\end{equation}
\rev{and,}
\begin{equation}
    \rev{\beta_{d2d} \in N(\beta,\sigma_{\beta}^2) }
\end{equation}
\rev{where $\sigma_{V_\alpha}$ and $\sigma_\beta$ are the standard deviation of the parameter $V_\alpha$ and $\beta$ in Eq. (6) and Eq. (7), respectively. Supplementary Fig. S9 shows the simulated conductance traces the program and erase cyclings as a function of program and erase times. The statistical results of the program time and erase time by the simulation are also shown in Fig. \ref{fig5}c and Fig. \ref{fig5}d, respectively.}

The high yield guaranteed by the production CMOS fabrication flow and high uniformity of analog switching demonstrated here promise the large-scale implementation of neuromorphic computations. However, learning from recent memristive neuromorphic research, we predict that two issues are limiting the performance of large scale neuromorphic systems based on Y-Flash memristor devices: (i) the non-ohmic readout behavior (note that the read currents in Fig. 2a are exponentially increasing with the applied voltage, thus the conductance is not constant); (ii) non-linear weight updates for the identical program and erase pulses (the read currents at 2V or conductance at 2V are exponentially dependent on the program/erase time in Fig. 3d and 3i). 

\rev{The non-ohmic behavior prevents the direct implementation of VMM on the Y-Flash memristor array shown in Fig. 1d for analogy input voltage vector. This issue can be solved by representing the analog input in the pulse width or number of pulses with a fixed read voltage\cite{Yao2017}, or by conducting the VMM in a small signal domain\cite{Danial2019}. In another aspect, in binary neural networks, with the input voltage vector being binarized\cite{Hirtzlin2020,Toledo2019}, the non-ohmic reading issue would not exist.}

\rev{The nonlinear weight update issue prevents precise weight tuning and is the major source of accuracy loss for online training of the mainstream deep neural networks\cite{Burr2015}. This issue can be solved by a write-and-verify-read (close-loop write)\cite{Yao2020} method, where multiple write operations are needed to tune the conductance to the targeted value with verifying readout operations. An alternative method is to utilize hybrid synaptic cells, incorporating a linear volatile part (for instance, a capacitor) to linearly update weights for each epoch and periodically transfer the volatile weights to the memristor part\cite{Ambrogio2018}. }

\rev{Novel solutions from the perspective of neural network structures and learning algorithms are needed to more efficiently use the Y-Flash memristor in a practical neuromorphic system. The success of these solutions needs precise simulations accounting for the exact electrical behavior of the Y-Flash memristors. The model developed here enables such precise simulations. } 

In summary, we develop an accurate compact model for the Y-Flash memristor. The model can work in various operational configurations since the missing voltage at floating terminals, including the FG, can be determined by analyzing the capacitor net. The I-V characteristics of the Y-Flash memristor can be accurately reproduced in both subthreshold and above-threshold regions. The switching behaviors (program and erase with voltage pulses) are modeled using equations that account for hot electron injection and band-to-band hot hole injection into the FG. The simulated  memristors are fabricated in the mass production CMOS process flows and demonstrate excellent reproducibility of parameters, thus making them promising for various neuromorphic applications.

\section*{SUPPLEMENTARY MATERIAL}
See supplementary material for more experimental data, modeling results, and the source code of the model.

\begin{acknowledgments}
This work was supported by the European Research Council through the European Union's Horizon 2020 Research and Innovation Programe under Grant 757259. W. Wang was supported in part at the Technion by the Aly Kaufman Fellowship.
\end{acknowledgments}

\section*{Data Availability Statement}
The data that support the findings of this study are available from the corresponding author upon reasonable request.

\section*{Conflict of interest}
The authors have no conflicts to disclose.

\bibliography{yflash_model}

\end{document}